\documentclass[journal]{IEEEtran}

\usepackage{silence}
\usepackage{xcolor,framed} 
\usepackage{soul}
\usepackage[cmex10]{amsmath}
\usepackage{amsfonts,amssymb}
\usepackage{bm}
\usepackage{color}  

\usepackage{bbding}
\usepackage{float}  
\usepackage{subfig}
\usepackage{graphicx}
\usepackage{caption} 

\usepackage{makecell}
\usepackage{textcomp}
\usepackage{array} 
\usepackage{longtable}
\usepackage{booktabs}
\usepackage{float}
\usepackage{threeparttable}
\usepackage{multirow}

\usepackage[numbers,sort&compress]{natbib}
\usepackage[hyphens]{url}
\usepackage[breaklinks, colorlinks, linkcolor=purple, anchorcolor=purple, citecolor=purple, urlcolor=purple]{hyperref}

\usepackage[ruled,norelsize,lined,linesnumbered]{algorithm2e}


\newtheorem{remark}{Remark}

\newtheorem{assumption}{Assumption}

\begin{document}
\title{Bidding and Dispatch Strategies with Flexibility Quantification and Pricing for Electric Vehicle Aggregator in Joint Energy-Regulation Market}
\author{Manqi~Xu,~\IEEEmembership{Student Member,~IEEE},
      Ye~Guo,~\IEEEmembership{Senior Member,~IEEE} and 
      Hongbin~Sun,~\IEEEmembership{Fellow,~IEEE.}
      \\
  \thanks{This work was supported by the National Natural Science Foundation of China under Grant 52377105. (Corresponding author: Y. Guo.)}
  \thanks{M. Xu and Y. Guo are with Tsinghua-Berkeley Shenzhen Institute, Shenzhen International Graduate School, Tsinghua University, Shenzhen, 518071, China. (e-mail: xmq22@tsinghua.mails.edu.cn; guo-ye@sz.tsinghua.edu.cn) }
  \thanks{Hongbin Sun is with the Department of Electrical Engineering, Tsinghua University, Beijing 100084, China (e-mail: shb@tsinghua.edu.cn).}
  }


\IEEEoverridecommandlockouts\IEEEaftertitletext{\vspace{-2.5\baselineskip}}
\maketitle

\begin{abstract}
Managing and unlocking the flexibility hidden in electric vehicles (EVs) has emerged as a critical yet challenging task towards low-carbon power and energy systems. 
This paper focuses on the online bidding and dispatch strategies for an EV aggregator (EVA) in a joint energy-regulation market, considering EVs' flexibility contributions and compensations.
A method for quantifying EV flexibility as a tradable commodity is proposed, allowing the EVA to set flexibility prices based on bid-in supply curves.
An EVA bidding model in the joint market incorporate flexibility procurement is formulated. The stochastic model predictive control technique is employed to solve the bidding problem online and address the uncertainties from the electricity markets and the EVs.
A power dispatch protocol that ensures a profitable and feasible allocation based on EV flexibility contribution is proposed. 
An affine mapping control strategy can be derived based on parametric linear programming, enables online indexing of optimal solutions given the regulation signals to avoid repeatedly solving the problem.
Numerical experiments show the effectiveness of the proposed scheme, and the solution methodology can be applied in real-time.

\end{abstract}

\begin{IEEEkeywords}
real-time bidding, power dispatch, electric vehicles, frequency regulation market, model predictive control, parametric programming
\end{IEEEkeywords}

%
\IEEEpeerreviewmaketitle


\vspace{-10pt}
\section{Introduction}
\subsection{Motivation}
\IEEEPARstart{E}{fforts} to decarbonize the power system and electrify transportation are driving the growth of renewable energy sources (RESs) and electric vehicles (EVs) \cite{res,iea}.
However, this growth creates a paradox:
The integration of uncoordinated charging load, coupled with the uncertain and volatile nature of RESs, can exacerbate power fluctuations and voltage deviation, threatening the power system reliability \cite{weiwei}.
Conversely, the elastic nature of EVs with the advancements in Vehicle-to-Grid (V2G) technology offers promising solutions to mitigate these negative impacts through coordinated EV charging \cite{v2g}.\par
A key question is how to effectively coordinate the rising number of EVs to enhance power system reliability.
EV flexibility can be leveraged by shifting charging demand or discharging within their sojourn times according to power system needs.
Direct control of a fleet of EVs by the independent system operator (ISO) is impractical due to the market participation thresholds and privacy concerns. 
This challenge has led to the emergence of the EV aggregator (EVA), which serves as an intermediary between EVs and ISO.

The profit-driven EVA typically participates in energy arbitrage and is increasingly involved in ancillary markets, such as reserve \cite{reserve1,reserve2}, frequency regulation \cite{allocation3,zhanghongcai_regulation,cvar} and demand response \cite{flex3,manqi} markets. This shift has created a win-win solution, enhancing both EVA profitability and power system reliability.
The frequency regulation market stands out as particularly lucrative among these ancillary markets. 
Hence, a fair and efficient EVA energy-regulation joint bidding model that utilizes EV flexibility and the subsequent power dispatch model in response to regulation signals are of great importance.

\vspace{-10pt}
\subsection{Related Works}
This section explores existing literature on three key aspects relevant to EVA operations: EV flexibility quantification and pricing, the incorporation of flexibility procurement in EVA bidding within the joint energy-regulation market, and the dispatch of EVs in response to regulation signals.\par

Quantifying EV users' flexibility contribution and compensating them appropriately is crucial. Several studies have explored demand-side flexibility quantification, generally categorizing approaches into region-based \cite{region1,flex1} and metric-based approaches \cite{flexibilityquantification, chenyue_p2p}.
The region-based method generally uses the power-energy feasible region of the EV charging problem.
Authors in \cite{flex1} proposed an optimization problem to characterize flexibility region as the area between upper and lower power trajectories. 
However, this region-based method describes the flexibility potential rather than flexibility contribution, which overlooks the actual flexibility that EVAs could harness from EVs, and market income is not allocated based on the actual flexibility contribution of EVs. 
Metric-based methods, on the other hand, measure flexibility in terms of capacity or power envelopes. 
Authors in \cite{flexibilityquantification} proposed three indices to quantify EV users' flexibility contribution based on upper energy boundary and actual energy level. While these indices provide a framework for understanding flexibility, they often restrict EVs from fully offering their potential and do not account for regulation contributions.

For the EVA bidding problem, one challenges is how to manage uncertainties in electricity markets and EV behaviors.
Typical techniques to manage uncertainties include stochastic optimization \cite{allocation3,stoc2,manqi,reserve1}, robust optimization \cite{robust,zhanghongcai_regulation,reserve2}, and risk-averse method \cite{zhanghongcai_regulation,cvar,stoc2}. 
However, these approaches lack adaptability, as they do not update in real-time, resulting in a static response to dynamic market conditions.
In contrast, the model predictive control (MPC) technique can address problems online and provides a certain degree of robustness to uncertainties through its receding-horizon implementation as in \cite{mpc1}. Nevertheless, its deterministic formulation typically limits its effectiveness in systematically managing uncertainties \cite{smpc}.
Besides, due to the EVA's intrinsic for-profit nature, the bidding strategies mentioned above \cite{allocation3,flex3,stoc2,zhanghongcai_regulation,manqi,reserve1,reserve2,robust,cvar,mpc1} aim to maximize profits across multiple electricity markets, benefiting both the EVA and the power system. 
However, these strategies often overlook the interests of EV users, which can lead to their resistance to market participation. Moreover, the flexibility procurement and bidding problems are inherently intertwined. Thus, how to quantify EV users' flexibility contribution and compensate them accordingly within the bidding problem remains an open question.

Besides the bidding problem, the subsequent power dispatch problem to determine the (dis)charging power of each connected EV within the EVA, is non-trivial and is not well studied, especially in the context of the regulation market. 
One challenge is how to ensure the fairness and feasibility of power allocation for each EV given the regulation signals. 
Proportional allocation \cite{fuzzy,reserve2} is the most common allocation protocol. Authors in \cite{reserve2} proposed a polytope-based EV aggregation and disaggregation method to bid in the energy and reserve market. 
Authors in \cite{allocation3} used the Lagrange multipliers of the energy update constraints in the bidding problem as the allocation protocol, ensuring that the EVA's profit maximization is still guaranteed.
However, these methods either overlook EVA’s profit maximization or EV users’ preferences and fairness.
Another significant challenge is adhering to the regulation signals issued by the ISO every few seconds.
Authors in \cite{allocation3,fuzzy} formulated the dispatch problem as a Mixed-Integer Linear Programming problem (MILP), solved each time a signal was issued. 
However, this method is computationally expensive and impractical, as solving the MILP can take several seconds.

\vspace{-10pt}

\subsection{Contributions and Organizations}
To address the aforementioned issues, this paper proposes online energy-regulation joint bidding and power dispatch models for EVA, incorporating EV flexibility quantification and pricing.
Our key contributions are as follows:\par
1) An EV flexibility quantification method that considers regulation provisions is proposed. By treating EV flexibility as a commodity, the EVA can set flexibility prices and procure them based on the bid-in flexibility supply curves of EVs.

2) An EVA bidding model in the joint energy-regulation market that incorporates flexibility procurement is proposed, aiming to minimize the EVA's total cost while respecting EVs' charging demand. The bidding problem is then reformulated using stochastic MPC technique, enabling real-time bidding.\par

3) An optimal power dispatch protocol that ensures EVA profitability while respecting EVs’ flexibility preferences and committed quantities is proposed. An affine mapping control strategy for the dispatch problem can be derived using parametric linear programming (pLP) prior to the operating hour, allowing for online indexing of optimal solutions every time a regulation signal is issued. 

The remainder of the paper is organized as follows: Section \ref{sOverview} provides an overview of the problem and market mechanisms. Section \ref{s3} details the quantification and pricing of EV flexibility. Section \ref{s4} describes the proposed EVA bidding and dispatch models. Section \ref{s5} explains the model reformulation and solution methodology. Section \ref{s6} presents simulation results, and Section \ref{s7} concludes the paper.
\vspace{-3pt}
\section{Overview}\label{sOverview}
This section first introduces the framework of our proposed EVA bidding and dispatch problems, and then presents the joint energy-regulation market mechanism.

\vspace{-5pt}

\subsection{Problem Overview}
As shown in Fig.\ref{FIG1}, the EVA bidding and dispatch problem adopts a three-layer structure, includes an ISO at the top layer, an EVA at the middle layer, and EVs at the bottom layer. 
As an intermediary, the tasks of the EVA mainly include two aspects:
\begin{figure}[h]
    \vspace{-10pt}
  \begin{center}
  \includegraphics[width=0.45\textwidth]{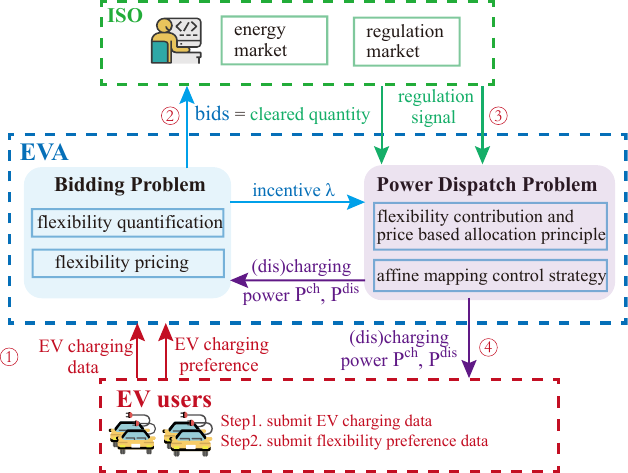}\\
  \caption{Overview of EVA bidding and dispatch problem.}\label{FIG1}
  \end{center}
  \vspace{-20pt}
\end{figure}

\textbf{Aggregating EVs and submitting bids to the ISO: }
\normalsize{\textcircled{\scriptsize{1}}} Each EV should share its charging demand and preference in offering flexibility to its connected EVA, who will conduct quantification and pricing of flexibility. 
\normalsize{\textcircled{\scriptsize{2}}} This procedure is embedded into the bidding problem, and EVA submits its joint energy-regulation offer/bid determined by the bidding problem to the wholesale market.

\textbf{Following regulation signals and dispatching power to EVs:}
\normalsize{\textcircled{\scriptsize{3}}} The ISO clears the offers/bids from all market participants and sends the cleared quantity back to the EVA.
\normalsize{\textcircled{\scriptsize{4}}} The EVA determines the dispatch power of each connected EV given the cleared quantity and the regulation signals.
\vspace{-10pt}
\subsection{Market Overview}

The market setting 
is based on the PJM market \cite{pjm}.
The energy market manages the sale and procurement of electricity to meet real-time and near-term demand.
Frequency regulation service is procured by the ISO to address frequency deviations caused by short-term electricity supply and demand mismatch. 
In PJM, upward and downward regulation are treated as the same product. 
meaning that resource owners who bid in the regulation market must be capable of delivering the bid-in quantity  $R_t$ in both directions. 
The regulation signal\footnote{PJM employs two types of regulation signals: the conventional RegA for resources with limited ramping capabilities and the dynamic RegD for resources with faster ramping but constrained energy capacity. Given the rapid ramping characteristics of EVs, our focus is on RegD signals.},
denoted as $s_{t,d}\in [-1,1]$, represents the ratio of a resource owner’s power change (upward or downward) to $R_t$ during sub-hourly interval $d$ within time $t:=\{1,....,D\}$. For example, in PJM, the regulation signals are issued every two seconds ($\Delta d =2s$), resulting in $D=1800$.
Following the regulation signal, the actual power of the resource owner is  $P^{act}_{t,d}=P^0_t-s_{t,d}R_{t}$.

A pay-for-performance mechanism is employed in the PJM regulation market, the resource owners will receive the following credit for providing regulations:
\begin{align}
    \Pi_t^{R}&=(c^{cap}_t+c^{per}_t m_t)R_t\label{reg_pay},
\end{align}
where $c^{cap}_t$ and $c^{per}_t$ are the hourly frequency capacity and performance clearing prices respectively, and $m_t$ is the regulation mileage at time $t$, represents accumulative absolute deviation of the adjacent regulation signals, $m_t=\sum_{d=1}^{D}|s_{t,d}-s_{t,d-1}|$.

The EVA energy and regulation offerings are interconnected, the actual power should not exceed the power limits $\underline{P}_t,\Bar{P}_t$ as follows. Consequently, the EVA jointly schedules its power profile and regulation capacities. 
\begin{align}
\underline{P}_t\leq P^{act}_{t,d}-R_t, \quad P^{act}_{t,d}+R_t\leq \Bar{P}_t.\label{cons2}
\end{align}

\begin{figure}[h]
  \begin{center}
  \includegraphics[width=0.4\textwidth]{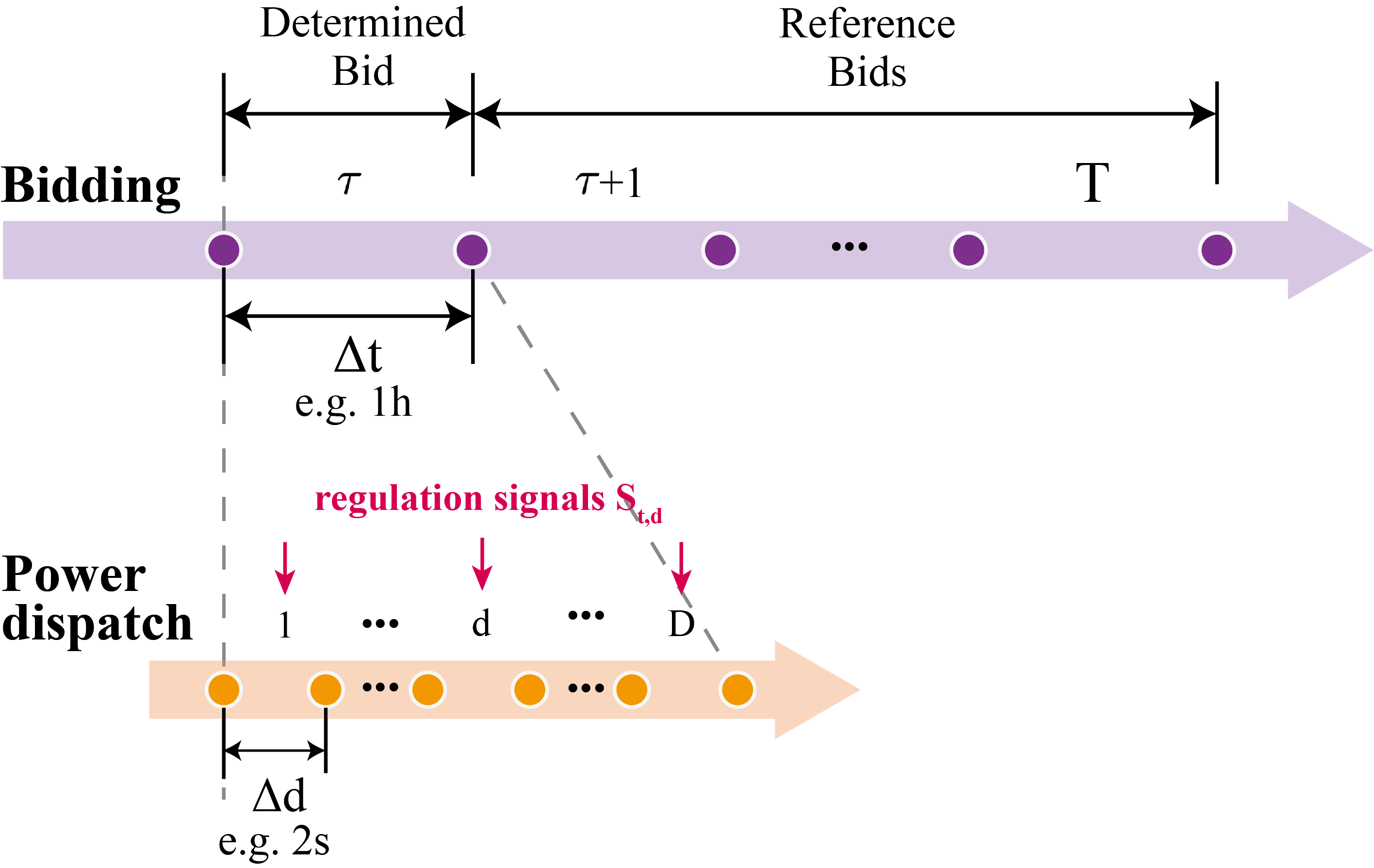}\\
  \caption{Real-time bidding and dispatch problem sequence.}\label{market sequence}
  \end{center}
  \vspace{-10pt}
\end{figure}

Additionally, since adjustments to regulation bids are allowed before the operating hour without any penalties, we focus on real-time bidding and dispatch strategies for the EVA. The sequence of theses processed is illustrated in Fig. \ref{market sequence}.

In each time interval $t$, the EVA determines the optimal bids based on market conditions and EVs' charging demands. After market clearing, the EVA will calculates dispatch strategies for all possible value of $s_d$ offline, and index for optimal dispatch solutions online given the regulation signals.

Compared to existing works \cite{zhanghongcai_regulation,allocation3}, the underlying features of the framework above are:
the integration of EV flexibility quantification and pricing into the bidding problem, and 2) the implementation of online dispatch given the regulation signals.
\vspace{-10pt}
\section{EV flexibility quantification and pricing}\label{s3}
In this section, we first derive the EV power and energy boundaries, and then present the EV flexibility quantification and pricing model.
\vspace{-10pt}
\subsection{Power and Energy Boundaries of Individual EVs}\label{Single EV Model}

The EVA can determine the power and energy boundaries based on the charging demand data shared by EV users, which includes $t_n^{a},  t_n^{d}, S_n^{a}, S_n^{d}, S_n^{min}, S_n^{max}, B_n$. 
Here,
$t^a$ and $t^d$ represent the arrival and departure times, while
$S_n^a, S_n^d, S_n^{min}$, and $ S_n^{max}$ denote the arrival, required, minimum and maximum state of charge (SoC) of the $n$th EV. $B_n$ represents the battery capacity. 
Before detailing the EV charging model, we outline the following assumptions for EVs: \par
\begin{assumption}
    The EV's sojourn time is defined as $[t^{a}_n,t^d_n)$, where $t^a$ and $t^d$ should be rounded for standardization.
\end{assumption}
\begin{assumption}\label{ass2}
    The expected SoC of the EV at departure will not exceed the SoC level achieved by charging at the maximum rate during its entire sojourn time, ensuring that the charging demand is feasible. 
\end{assumption}
\begin{assumption}
    The connected chargers support V2G mode, allowing the EV to choose whether to charge or discharge.
\end{assumption}
Let $\mathbf{T}=\{1,2,\dots,T\}$ represent the set of time intervals. 
For the $n$th EV, its charging power $p_{n,t}^{c}$, discharging power $p_{n,t}^{d}$ and energy level $e_{n,t}$ must satisfy the following constraints:
\begin{subequations}\label{bc}
\begin{align}
 & p_{n}^{min}\leq p_{n,t}^{c}- p_{n,t}^{d}\leq p_{n}^{max}, \forall t \in [t_n^{a},t_n^{d}), \label{1power_boubd}\\
 & 0\leq p_{n,t}^{c} \perp p_{n,t}^{d} \geq 0, \forall t \in [t_n^{a},t_n^{d}),\label{1non_neg}\\
 & p_{n,t}^c=0,p_{n,t}^d=0, \forall t\notin [t_n^{a},t_n^{d}),\label{1pnot_connected}\\
 & e_{n,t+1}=e_{n,t}+p_{n,t}^c\eta^c\Delta t-p_{n,t}^d \Delta t/\eta^d, \forall t\neq T ,\label{1energy_update}\\
 & e_{n,t}=e_n^a,  \forall t\in[1,t_n^{a}],e_{n,t}\geq  e^d_t ,  \forall t\in[t_n^{d},T],\label{1ed}\\
 & e_{n}^{min}\leq e_{n,t} \leq e_{n}^{max}, \forall t\in \mathbf{T}. \label{1energy_boun}\end{align}
\end{subequations}

\textbf{EV power constraints (\ref{1power_boubd})-(\ref{1pnot_connected}):} These constraints limit the (dis)charging power within its minimum and maximum rated power, $p_{n}^{min}$ and $p_{n}^{max}$, during its sojourn time. Additionally, the charging and discharging power are non-negative and cannot occur simultaneously.

\textbf{EV energy constraints  (\ref{1energy_update})-(\ref{1energy_boun}):} 
Constraint (\ref{1energy_update}) updates the cumulative EV energy, with $\eta^c$ and $\eta^d$ as charging and discharging efficiencies. 
Constraint (\ref{1ed}) ensures the fulfillment of charging demand, from the initial energy level $e^a_n=B_n S_n^a$ to the desired energy level $e^d_n=B_n S_n^d$, while constraint (\ref{1energy_boun}) maintains the energy within the safe bounds, defines by $e^{min}_n=B_n S_n^{min}$ and $e^{max}_n=B_n S_n^{max}$. \par

The power and energy boundaries can be derived from constraints (\ref{bc}), which define the envelopes of the feasible region for net power $p^c_{n,t}-p^d_{n,t}$ and energy $e_{n,t}$. Note that $e_{n,t}^+$ reflects a strategy to charge to $S^{max}$ as quickly as possible, while the opposite is true for $e_{n,t}^-$ .

\begin{subequations}
\vspace{-10pt}
\begin{align}
& p_{n,t}^{+/-}=\left\{\begin{array}{l}
p_{n}^{max/min}, t_n^{a} \leq t < t^{d}_n,\\
0, else,
\end{array}\right. \label{power upper bound}\\
& e_{n, t}^+=\left\{\begin{array}{l}
e^{a}_n, t\leq t_n^{a},\\
e^{max}_n, t>t^{d}_n,\\
\min \left\{e^{a}_n+\eta_n^{c} p_{n}^{max}\left(t-t^{a}_n\right),e^{max}_n\right\},else,
\end{array}\right.\\
& e_{n, t}^-=\left\{\begin{array}{l}
e^{a}_n, t\leq t_n^{a},\\
e^{d}_n, t>t^{d}_n,\\
\max \{e^{a}_n, e^{d}_n+\frac{p_{n}^{min}(t^{d}_n-t)}{\eta_n^{d}}, e^{min}_n\},else.
\end{array}\right. \label{energy lower bound}
\end{align}
\end{subequations}

\subsection{EV flexibility quantification}\label{sec_flex}
EV flexibility is promised through allowances in various charging trajectories within the predefined power and energy boundaries. 
While EVs' deviations from the baseline power profile benefit the EVA in the electricity market, they may reduce the satisfaction of EV users. Therefore, EV users should be compensated accordingly.
If Assumption \ref{ass2} holds, the EVA can satisfy the charging demand of EVs before its declared departure time, and EVs will not be redundant to charging load shifting.
However, discharging power will adversely affect the EV battery, and both upward and downward regulation ranges should be accounted for and compensated. 
We define the baseline power ${p}^0_{n, t}$ as the EV power profile when the EVA bids in the joint energy-regulation market, and define the overall flexibility contribution of the $n$th EV at time $t$ as:
\begin{alignat}{2}
    & Flex_{n,t} \triangleq (  \underbrace{p_{n,t}^{d}/\eta_n^{d} }_\text{degradation}+\underbrace{\Delta p_{n,t}^{up}+\Delta p_{n,t}^{dn} }_\text{upward/downward regulation})\Delta t\label{7-flex_sum}.
\end{alignat}

The first term reflects the actual discharging power of the $n$th EV at time $t$, defined as the discharging power $p^d_{n,t}$ adjusted by the discharging efficiency. Its relationship with $p^0_{n,t}$ is given by $p^0_{n,t}=p^c_{n,t}-p^d_{n,t}$. This term is crucial, as EV users generally disfavor discharging due to its impact on battery degradation.
Many prior studies \cite{reserve1,allocation3,zhanghongcai_regulation,cvar} have employed a similar formulation in bidding models to compute degradation costs, which partially consider the interests of EV users, but do not fully capture the trade-off that EVs make for optimal coordinated charging under EVA's direction.\par

The second term represents the upward and downward regulation capacity contributions of the $n$th EV at time $t$, denoted as $\Delta p^{up}_{n,t}$ and $\Delta p^{dn}_{n,t}$, respectively. These contributions are constrained by the power and energy boundaries.

Typically, the regulation capacity of a resource owner is tied to $p^0_{n,t}$ and limited solely by the power boundaries, which can be readily derived from constraint (\ref{cons2}) as follows: 
\begin{subequations}
\begin{align}
    & 0\leq \Delta p^{up}_{n,t}\leq p^0_{n,t}-{p}^-_{n,t}, \quad 0\leq \Delta p^{dn}_{n,t}\leq {p}^+_{n,t}-p^0_{n,t}.\label{4pdn}
\end{align}

However, EVs must also meet their charging demands as follows, which makes their baseline power time-dependent,
   \begin{align}
    & p^0_{n,t}-\Delta p^{up}_{n,t}=p^{c,up}_{n,t}+p^{d,up}_{n,t},\label{4c}\\
    & p^0_{n,t}+\Delta p^{dn}_{n,t}=p^{c,dn}_{n,t}+p^{d,dn}_{n,t},\label{4d}\\
    & e_{n,t+1}^{\text{up/dn}}=e_{n,t}^{\text{up/dn}}+p^{c,\text{up/dn}}_{n,t}\eta^c\Delta t-\frac{p^{d,\text{up/dn}}_{n,t}\Delta t}{\eta^d},\label{4e}\\
    & {e}_{n,t}^-\leq e_{n,t}^{up/dn}\leq {e}_{n,t}^+.\label{4f}
\end{align} 
\end{subequations}

Constraints (\ref{4c}) and (\ref{4d}) specify the actual charging and discharging power of the $n$th EV at time $t$ when its committed upward and downward reserve capacities, $p^{c,up/dn}_{n,t}$ and $p^{d,up/dn}_{n,t}$, are fully deployed, respectively.
Constraint (\ref{4e}) governs the energy update, while constraint (\ref{4f}) ensures that the EV's charging needs can be satisfied in boundary conditions.

\begin{figure}[!ht]
    \vspace{-10pt}
    \centering
    \includegraphics[width=0.4\textwidth]{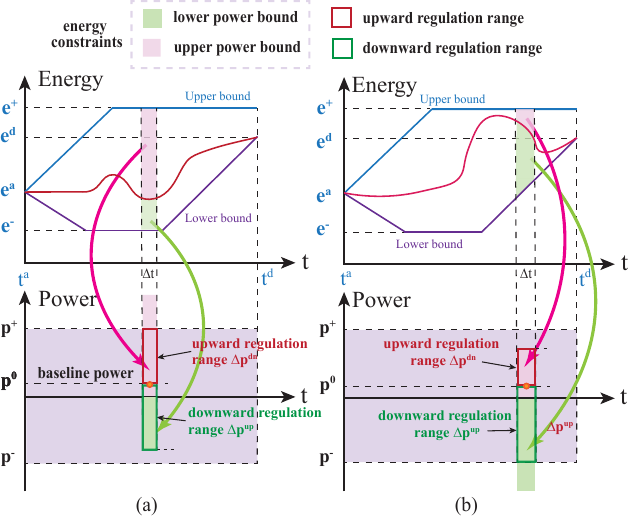}
    \captionsetup{font={footnotesize}, justification=raggedright}
    \caption{Illustration of upward/downward reserve range: (a) the constraints on the right side of (\ref{4pdn}) and left side of (\ref{4f}) are binding; (b) constraints on the left side of (\ref{4pdn}) and right side of (\ref{4f}) are binding.}
    \label{fig:illus}
    \vspace{-10pt}
\end{figure}

We illustrated the upward and downward regulation capacities and their relationship with baseline power in Fig. \ref{fig:illus}.

\subsection{EV Flexibility Bid-in Supply Curve}
\begin{figure}[!ht]
    \centering
    \includegraphics[width=0.5\linewidth]{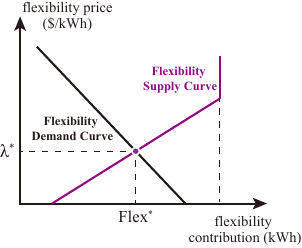}
    \caption{EV flexibility bid}
    \label{fig:enter-label}
    \vspace{-10pt}
\end{figure}
Our flexibility pricing method is based on the willingness of EV users to offer flexibility. Each EV user submits a supply curve to indicate their willingness, defined as follows:
\begin{align}
& Flex_{n,t} = k_{n,t}\lambda_{n,t} + \xi_n \label{supply curve}, 
\end{align}
where $Flex_{n,t}$ represents the amount of flexibility contributed by the EV defined in (\ref{7-flex_sum}), $\lambda_{n,t}$ is the flexibility price given by the EVA, $k_n$ denotes the EV flexibility offering preference, and $\xi_n$ represents the amount of flexibility the EV will offer without any compensation. 

The determination of $k_n$ is based on the maximum flexibility and EV charging fee:
\begin{align}
    & k_{n,t}\triangleq \alpha_n {Flex^{max}_{n}}/{c^{fee}_t}, \\
    & Flex^{max}_{n}\triangleq(-\frac{p^{min}_n}{\eta_d}+p^{max}_n-p^{min}_n)\Delta t,\label{flex_max}
\end{align}
where $\alpha_{n,t}$ is flexibility offering preference ratio,  $c^{fee}$ is EV charging price, and $Flex^{max}_{n}$ is the typical maximum flexibility contribution of the $n$th EV as defines in (\ref{flex_max}). 
A higher value of $k_n$ indicates that the EV is more elastic to the offered price. 
From the EV user's perspective, it only needs to submit $(\alpha_n,\xi_n)$ to indicate its charging preferences, which are key to determining its flexibility offering behavior.

\section{EVA Bidding Model with EV Flexibility Procurement}\label{s4}

This section we first introduces the EVA bidding strategy for optimal aggregated bids, incorporating flexibility procurement. Then we reformulate the bidding problem using the stochastic MPC technique to address uncertainties in market conditions and EV behaviors, enabling online solutions.

\subsection{Flexibility embedded EVA bidding problem}\label{biddingmodel}
We define $\mathcal{N}=\{1,...,N\}$ as the set of EVs in an EVA, each of which is indexed by $n$. 
Each EV user is required to submit the charging demand ($t_n^{a}, t_n^{d}, S_n^{a}, S_n^{d}, B_n$) and flexibility preference data $(\alpha_n,\xi_n)$ to the connected EVA.
Before detailing the model, we make the following assumption for the EVA bidding problem:
\begin{assumption}\label{ass4}
     The EVA is treated as a price-taker in the energy-regulation market joint bidding process as in \cite{reserve1,flex2}. Optimal quantity-only joint bids for energy and regulation are determined to minimize the expected cost.
\end{assumption}\par
The objective function and constraints of the proposed bidding problem are presented below.
\subsubsection{Objective function}

Besides the commonly considered EVA's market income, our proposed bidding strategy also considers the interests of EVs by incorporating EV flexibility procurement. 
The objective function includes EVA's total expected cost associated with both the ISO and the EVs. 
The EVA's net cost with the ISO at time $t$ is given by
\begin{alignat}{2}
   \Phi^{\textit{ISO}}_t =& c_t^{e} P_t^e-(c_t^{\mathrm{rc}}+c_t^{\mathrm{rp}} m_{t}) R_{t}\label{obj_bid1},
\end{alignat}
where the first term denotes the net energy cost, where $c^{e}_t$ is the predicted real-time energy market price at time $t$, $P_t^e$ is the bid-in energy quantity at time $t$, energy arbitrage can help reduce this cost. The second term denotes credit for offering frequency regulation service, as explained in (\ref{reg_pay}).

The EVA's net cost with $n$th EV at time $t$ is
\begin{alignat}{2}
   \Phi^{EV}_{n,t}=&
        \lambda_{n,t} Flex_{n,t} -c^{fee}_t p^0_{n,t},
\end{alignat}
where the first term denotes the payment to EV for its flexibility contribution, as explained in (\ref{7-flex_sum}) and (\ref{supply curve}), and second term denotes the EV charging income.

The EVA's revenue from providing grid services can be passed on to EV users and allocated among them based on flexibility contribution.
Note that for flexibility quantification, $p^d_{n,t}$, $\Delta p^{up}_{n,t}$ and $\Delta p^{dn}_{n,t}$ are deeply coupled with $p^0_{n,t}$  and are determined in the bidding problem. 
The flexibility price $\lambda$ set by the EVA is subsidized by the profits EVA earned from energy arbitraging and frequency regulation market. Therefore, flexibility payments are inherently intertwined with the bidding decisions and are integrated into the EVA's bidding problem for joint decision-making.

\subsubsection{Constraints}
In addition to the EV flexibility-related constraints outlined in equations (\ref{7-flex_sum})-(\ref{supply curve}), market bidding constraints are further specified as follows.
The EVA's energy bid-in quantity $P^e_t$ equals the aggregated baseline charging power $P^0_{n,t}$ scheduled for all EVs:
\begin{align}
    & P_{t}^e=\sum_{n\in N} p^0_{n,t}.\label{pe}
\end{align}

The EVA's frequency regulation bid-in quantity should be non-negative and is limited by both the aggregated upward and downward adjustment capabilities:
\begin{align}
    & 0\leq R_{t}\leq \min \{\sum_{n\in N}\Delta p_{n,t}^{up},\sum_{n\in N}\Delta p_{n,t}^{dn}\}. \label{2-regulation capacity bounds}
\end{align}

Finally, we can formulate the EVA’s multi-period co-optimization bidding model as:
\begin{subequations}\label{bidding_problem}
\begin{alignat}{2}
   \textbf{(P1)}\min_x & \quad \sum_{t\in \mathbf{T}} \sum_{n\in \mathcal{N}}\Phi^{\textbf{ISO}}_t+\Phi^{\textbf{EV}}_{n,t},\\
s.t.& \quad(\ref{7-flex_sum})-(\ref{supply curve}),(\ref{pe})-(\ref{2-regulation capacity bounds}),\forall  t \in \mathbf{T},\forall n \in N,\label{8k}
\end{alignat}
\end{subequations}
where $x:=\{P_t^e$$, R_t, \Delta p^{up/dn}_{n,t}, $$p^0_{n,t}, e_{n,t}^0, e_{n,t}^{up/dn},\lambda_{n,t}, $ $Flex_{n,t}$$|\forall t \in \mathbf{T},$ $\forall n$ $\in \mathcal{N}\}$ is the set of decision variables of the bidding problem. 

\begin{remark}
Although the PJM regulation market requires the resource owners to be capable of deploying bid-in capacity in both directions, as constrained in (\ref{2-regulation capacity bounds}), this requirement does not apply to individual EVs when bidding through an EVA, as described in (\ref{4pdn}). The EVA can fully leverage the upward and downward flexibility of the connected EVs, as long as their aggregated regulation bid-in quantity can be met in both directions. This EVA-based model enhances overall regulation capacity, lowers barriers for small resources to meet market thresholds, and improves regulation utilization efficiency.

\end{remark}
\vspace{-10pt}
\subsection{Stochastic MPC-Based Bidding Problem Reformulation}
We employ the stochastic MPC technique to integrate probabilistic descriptions of regulation signal into the \textbf{P1}.
This approach allows the EVA to dynamically adjust its bidding strategy over time by solving an open-loop constrained optimal control problem in a time-receding manner, ensuring timely decision-making before each operating hour.

Based on regulation signals, scenarios are categorized into typical and extreme sets, denoted as $\mathcal{A}$ and $\mathcal{B}$, respectively.  
$\mathcal{A}$ includes boundary values -1 and 1, for which we calculate the occurrence probabilities at each case.
$\mathcal{A}$ are defined by dividing the range into smaller and evenly distributed intervals, including boundary values -1 and 1. 
Note that typical scenarios $\mathcal{A}$ are used to consider the uncertain nature of frequency signals to ensure EVA profitability, while extreme scenarios $\mathcal{A}$ ensure the feasibility of power dispatch across all EVs. This approach considers both economic performance and operational reliability under uncertainty.
\subsubsection{Objective}
Given the current time slot $\tau$, the bidding horizon $\mathcal{T}:= \{\tau, \tau +1, ..., T\}$ encompasses the present and future periods.
\begin{subequations}
\begin{alignat}{2}
       \mathcal{J}=&  \sum_{t\in \mathcal{T}} \sum_{n\in \mathcal{N}} \sum_{\omega\in \Omega}
  \left (\begin{array}{l}
         \Phi^{\textbf{ISO}}_t+\Phi^{\textbf{EV}}_{n,t}\\
        +\pi_{\omega,t}c^{dp}_t (\delta p^{up}_{\omega,n,t}-\delta p^{dn}_{\omega,n,t})
   \end{array}
   \right),\label{obj_bid2}
\end{alignat}
where the new term is the expectation of re-dispatch costs in all scenarios,  $\delta p^{up}_{\omega,n,t}$ and $\delta p^{dn}_{\omega,n,t}$ denote the power upward and downward adjustments, respectively.
\subsubsection{Constraints}
In addition to the EV flexibility and market bidding constraints, the reformulated stochastic MPC-based bidding problem is also constrained by EV power and EV energy constraints for all scenarios.
The EVs' power constraints for all $\omega \in \mathcal{A} \bigcup \mathcal{B}$ include:
\begin{align}
    & P_{t}^e-s_{\omega,t} R_{t}=\sum_{n\in \mathcal{N}}(p^0_{n,t}-\delta p^{up}_{\omega,n,t}+\delta p^{dn}_{\omega,n,t}), \label{1-balance}\\
& p_{\omega,n, t}^{c}-p_{\omega,n, t}^{d}=p^0_{n,t}-\delta p^{up}_{\omega,n,t}+\delta p^{dn}_{\omega,n,t},\label{7f}\\
& 0 \leq p_{\omega,n, t}^{c}\leq (1-\mu_{\omega,n,t}){p}^+_{n, t}, \label{3_non}\\
& 0 \leq p_{\omega,n, t}^{d}\leq \mu_{\omega,n,t}{p}^-_{n, t},\label{4-dis_non}\\
& 0 \leq \delta p_{\omega,n, t}^{up/dn}\leq \Delta p_{n, t}^{up/dn}.\label{8h}
\end{align}

Constraint (\ref{1-balance}) ensures that EVA will follow the regulation signals $s_{\omega,t}$ in all scenarios by adjusting the baseline power $p^0_{n,t}$ of the connected EVs.
Constraint (\ref{7f}) defines the relationship between the EV's actual (dis)charging rate and its baseline power, adjusted by the upward or downward regulation.
Constraints (\ref{3_non}) and (\ref{4-dis_non}) introduce binary variables $\mu_{\omega,n,t}$ to eliminate the bi-linear terms in (\ref{1non_neg}). They ensure that both charging and discharging power remain non-negative, simultaneous charging and discharging is prohibited, and the charging and discharging power are confined within the EV's allowable power boundaries.
Constraint (\ref{8h}) ensures that the EV's upward and downward power adjustments to the baseline power remain within the committed regulation range $\Delta p^{up/dn}_{n,t}$. 

The EVs' energy constraints include:

\begin{alignat}{2}
& e_{n, t+1}^0=e_{n,t}^0+ \sum_{\omega \in \mathcal{A}}
\pi_{\omega} \left(p_{\omega, n,t}^{c}\eta^{\mathrm{c}}-\frac{p_{\omega, n, t}^{d} }{\eta^{d}} \right)\Delta t, \label{6bid-energy update1}\\
& {e}^-_{n, t} \leq e_{n, t}^0 \leq {e}^+_{n, t}.\label{7bid-energy boundary1}
\end{alignat}
\end{subequations}

The energy update for EVs in typical scenarios is given in constraint (\ref{6bid-energy update1}), while constraint (\ref{7bid-energy boundary1}) ensures that the cumulative energy remains within the upper and lower limits. 

Overall, the EVA can make real-time bidding decisions by solving a rolling-window stochastic optimization problem as follows:
\begin{align}
    (\textbf{P1'})\min_{x_\tau} & \quad \mathcal{J},\\
    s.t. & \quad(\ref{7-flex_sum})-(\ref{supply curve}),(\ref{pe})-(\ref{2-regulation capacity bounds}),\forall  t \in \mathcal{T},\forall n \in \mathcal{N},\\
    & \quad(\ref{1-balance})-(\ref{7bid-energy boundary1}),\forall \omega \in \mathcal{A},\forall  t \in \mathcal{T},\forall n \in \mathcal{N}.
\end{align}
where $x_{\tau}:=\{P_t^e, R_t, P^{c/d}_{\omega,n,t},\Delta P^{up/dn}_{n,t}, P^0_{n,t}, \delta P^{up/dn}_{\omega,n,t},E_{n,t}^0,$ $\lambda_{n,t},$ $Flex_{n,t}|\forall t \in \mathcal{T},\forall n\in \mathcal{N}\}$ is the set of decision variables of the reformulated bidding problem.

The EVA periodically solves \textbf{P1'} over the entire horizon $\mathcal{T}$, but only the solutions for the current time step ($\tau$), $P_{\tau}$ and $R_{\tau}$, are submitted to ISO as bid-in quantity. The remaining time slots $\mathcal{H}:= \{\tau+1, ..., T\}$ serve only as forecasts and are adjusted in subsequent periods.
And EVA will update the parameters $P_t$, $R_t$,$\lambda_{n,t}, P^0_{n,t}$ and $\Delta P^{up/dn}_{n,t}$ to the power dispatch problem as fixed parameters.

\begin{remark}
    The nonlinearity in the objective function, arising from the term $\lambda_{n,t}Flex_{n,t}=k_n\lambda_{n,t}^2+\xi_n \lambda_{n,t}$, makes it a convex mixed integer quadratic programming (MIQP). While MIQP can be solved efficiently using methods like branch-and-bound, extended cutting planes, and benders decomposition \cite{kronqvist_review_2019}, the detailed solution techniques are beyond the scope of this paper. 
\end{remark}

\section{EVA Power dispatch Model}\label{s5}
In this section, we discuss the power dispatch strategy given regulation signals and then reformulate it as a pLP problem to enable fast and efficient computation.

\subsection{EVA Power Dispatch Problem}
After market clearing, every time a regulation signal is received, the EVA must solve the dispatch problem to determine the dispatched power for all connected EVs.
This process presents two key challenges: 
\romannumeral1) Proper power dispatching: How to efficiently allocate the total power among the EVs in response to the regulation instructions, ensuring both feasibility and profitability?
\romannumeral2) Online dispatch: How to rapidly compute power allocation decisions upon receiving frequent regulation signals within short time intervals?

We first introduce the power dispatch model for the EVA, and the next subsection introduces how to solve the problem online. Note that we omitted the subscript $t$ thereafter.
\begin{assumption}
    We assume that EV SoC is not binding within each sub-hourly interval. 
Therefore, minor energy updates during sub-hourly intervals can be disregarded. 
\end{assumption}

\subsubsection{Objective function}
The allocation protocol is to minimize the cost for EV flexibility compensation and power adjustment based on flexibility bids from EVs, which is expressed as
{\small
\begin{alignat}{2}
    & F_d(p_{n,d}^{d},\delta p_{n,d}^{up},\delta p_{n,d}^{dn})\notag\\
    =& \sum_{n\in N} \lambda_{n} flex_{n,d}+c^{dp}_n(-\delta p^{up}_{n,d}+\delta p^{dn}_{n,d})\notag\\
     = &\sum_{n\in N}\left( \frac{\lambda_{n} }{\eta_n^{d}}p_{n,d}^{d} +(\lambda_n-c^{dp}_n)\delta p_{n,d}^{up}+(\lambda_n+c^{dp}_n)\delta p_{n,d}^{dn}\right).
\end{alignat}
}

Here, the flexibility prices $\lambda_{n}$ are determined in the bidding problem, reflecting both the EV users' willingness to provide flexibility and the scarcity of flexibility resources. EVs with lower $\lambda_{n}$ are prioritized to meet higher flexibility needs. 

This can also be viewed as a regulation command $s_{t,d}R_t$ allocation problem in which the EVA should determine the power adjustments $\delta p^{up/dn}_{n,t}$ for all connected EVs.
\subsubsection{Constraints}
The cleared quantities $\hat{P}_t$ and $\hat{R}_t$ are fixed parameters in the dispatch problem. Upon receiving a regulation signal $s_{d}$ from the ISO, the total power to be dispatched among the EVs, $\hat{P}_t-s_d \hat{R}_t$, is determined. 
Hence, ensuring feasibility can be framed as a resource allocation problem with respect to the operational limits of EVs, which must satisfy the following constraints:
\begin{subequations}
    \begin{alignat}{9}
    & \sum_{n\in N}(p_{n,d}^{c} - p_{n,d}^{d})=\hat{P}-s_d\hat{R},\label{balance_resource}\\
    & p_{n,d}^{c}-p_{n,d}^{d}=\hat{p}^0_{n}-\delta p_{n,d}^{up}+\delta p_{n,d}^{dn}, \forall n\in N ,\label{9b}\\
    & 0 \leq p_{n,d}^{c} \leq {p}^+_n, 0 \leq p_{n,d}^{d}\leq -{p}^-_n,\forall n\in N,\label{9c}\\
    & 0\leq \delta p_{n,d}^{up/dn}\leq \Delta p_{n,d}^{up/dn},\forall n\in N ,\label{9 devirange}\\
    &  p_{n,d}^{c} \perp p_{n,d}^{d},\delta p_{n,d}^{up}\perp \delta p_{n,d}^{dn},\forall n\in N .\label{9e}
\end{alignat}
\end{subequations}

Constraint (\ref{balance_resource}) ensures that the total wait-to-allocate power matches the sum of all power assigned to the EVs.
Constraint (\ref{9b}) defines the relationship between the actual charging/discharging power of each EV and its baseline power, adjusted by the upward or downward regulation.
Constraint (\ref{9c}) states the same as constraint(\ref{1non_neg}).
Constraint (\ref{9 devirange}) ensures that the actual upward and downward regulation flexibility is non-negative and within the predetermined regulation bounds $\Delta p_{n,d}^{up/dn}$ from the bidding process to make sure EV's charging demand can be met. 
Finally, constraint (\ref{9e}) prohibits simultaneous upward and downward regulation to prevent EV from flexibility arbitrage.

The overall single-period power dispatch problem can be formulated as follows, with the decision variables denoted as $x=\{p^{c/d}_{n,d},\delta p^{up/dn}_{n,d}|\forall d\in \mathcal{D},\forall n\in \mathcal{N}\}$:
\begin{alignat}{2}
    \textbf{P2:} \quad &\underset{x}{\operatorname{min}}\quad  F_d(\cdot),
    \quad s.t.\quad  (\ref{balance_resource})-(\ref{9e})\notag.
\end{alignat}

\subsection{Affine mapping control strategy of dispatch problem}\label{dispatch2}
$\mathbf{P2}$ is typically solved by the EVA upon receiving a regulation signal, but is computationally intractable in its current form for two reasons:
\romannumeral1) The bi-linear constraints introduce nonlinearity, complicating the solution process. 
\romannumeral2) The dynamic nature of regulation signals results in solution times that exceed the interval between regulation signal updates.

For the first issue, the bi-linear constraint $P^c_{n,t} \perp P^d_{n,t}$ can be exactly relaxed, as simultaneous charging and discharging provides no benefit—charging incurs no gain, while discharging imposes additional costs. Similarly, the bi-linear term $\delta P^u_{n,t} \perp \delta P^d_{n,t}$ can be removed when $\lambda_n - c^{dp}_n \leq 0$. This transforms the problem into a linear programming formulation.

For the second issue, we use parametric programming by treating $s_{t,d}$ as a parameter, adopting an affine mapping control strategy that enables the EVA to efficiently determine online dispatch based on the regulation signal.

Accordingly, by handling these two aspects, we reformulate the problem \textbf{P2} as:
\begin{subequations}\label{power_dispatch4}
\begin{alignat}{2} 
    (\textbf{P2'})\mathcal{V}(\bm{\theta})& = 
    \min_{\bm{x}}  \bm{c}^{\intercal} \bm{x},\label{P3-obj}\\
    s.t. &  \bm{A}^{eq}\bm{x} = \bm{b}^{eq}+\bm{F}^{eq} \bm{\theta}\label{2-eq}, \bm{A}\bm{x}\leq \bm{b},\\
  & \bm{A}^{t} \bm{\theta} \leq \bm{b}^t\label{2-theta},
\end{alignat}
\end{subequations}
where $\bm{x}$ and $\bm{\theta}$ are vectors of decision variables and parameter vector. \textbf{P2'} is subject to the equality and inequality constraints in (\ref{2-eq}) and parameter boundary constraint in (\ref{2-theta}). 


After market clearing and before the operating hour, the EVA can obtain the critical regions (CRs) along with the corresponding optimal solutions and value function.

The CR partition the parameter space $\Theta$ into intervals, where the set of active and inactive constraints remains unchanged.
Within each CR, the optimal solution $\bm{x}^*(\bm{\theta})$ and the value function $\mathcal{V}(\bm{\theta})$ can be expressed as affine functions of $\theta$:
\begin{align}
    & \bm{x}^*(\bm{\theta})=\bm{R}\bm{\theta}+{\bm{r}},\\
    & \mathcal{V}^*(\bm{\theta}):= 
        \bm{c}^\intercal \bm{R} \bm{\theta}+\bm{c}^\intercal \bm{r}
\end{align}
where $\bm{R}$ and ${\bm{r}}$ can be obtained through Karush-Kuhn-Tucker conditions under active and inactive sets, see \cite{gy}.

This proposed power dispatch strategy substantially reduces computational complexity by transforming the problem from one that requires solving it separately for each sub-interval $d$ to one that only needs to be solved once over the entire $t$.

\section{Case Study}\label{s6}
\subsection{Simulation Settings}
All simulations were done on a laptop with an Apple M3 Pro processor and 18 GB of memory. The bidding problems were programmed by Python and solved using Gurobi solver v11.0.3 \cite{gurobi}, and the dispatch problem was programmed by Matlab and solved using MPT3 \cite{MPT3}. 

To verify the effectiveness of the proposed methods, simulations are conducted using an EVA comprising 100 EVs. 
Each EV is assumed to have a battery capacity $B$ of 50 kWh. The SoC ranges were set to a maximum of 90$\%$ and a minimum of 20$\%$. 
Charging efficiency $\eta _c$ and discharging efficiency $\eta _d$ are assumed to be 90$\%$ and 93$\%$, respectively.
Monte Carlo method is used to generate EV users' behavior data, as outlined in Table \ref{tab:2}, conforming to a specified probability distribution\cite{gaussian}.
To verify the proposed flexibility quantification and pricing methods, we 
divide EVs into three types based on their response preferences: conservative, cautious, proactive and risky, with corresponding values $\alpha_{n}$ as 0.2, 0.6, 1, and 1.4. And set $\xi$ all zero.
\begin{table}[h]
	\centering
	\caption{Probability distribution of EV charging data}
	\label{tab:2}  
    \begin{threeparttable}
	\begin{tabular}{cccc cc}
		\hline\hline\noalign{\smallskip}	
		Parameters & Distr.  & $\mu$ & $\sigma$& $Min$& $Max$   \\
		\noalign{\smallskip}\hline\noalign{\smallskip}
		Arrival Time (h)  & TND & 18 & 1 & 13 & 24 \\
		Departure Time (h)& TND & 8 & 2 & 1 & 12 \\
            Arrival SoC ($\%$) & UD & 30 & - & 20 & 40 \\
            Required SoC ($\%$) & UD & 80 & - & 70 & 90 \\
            Max. (dis)charging power (kW) & UD & 10  & - & 8 & 12 \\
		\noalign{\smallskip}\hline
	\end{tabular}
     \begin{tablenotes}
            \footnotesize
            \item * TNG: Truncated Normal Distribution, UD: Uniform Distribution.
          \end{tablenotes}
      \end{threeparttable}
      \vspace{-10pt}
\end{table}

\begin{figure}[ht]
  \begin{center}
  \includegraphics[width=0.4\textwidth]{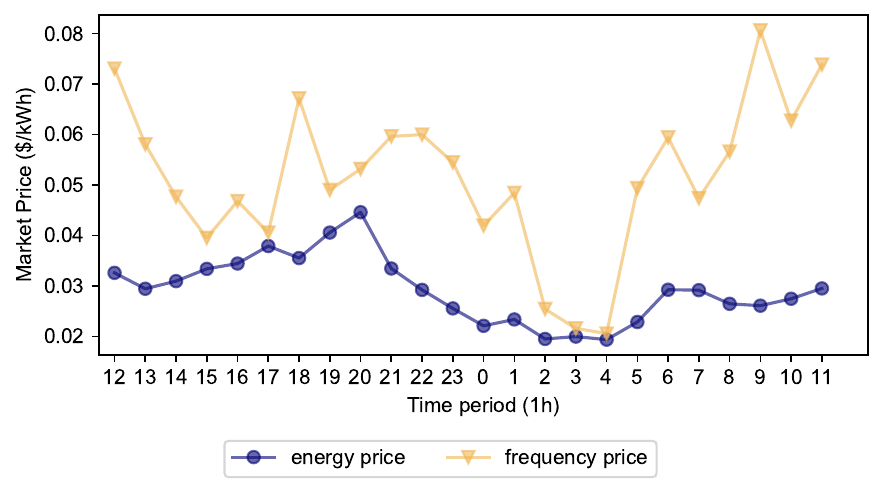}\\
  \caption{Market price.} \label{price_fig}
  \end{center}
  \vspace{-15pt}
\end{figure}

The entire simulation duration covers 24 hours. The market prices are shown in Fig. \ref{price_fig}, where
we leveraged historical data from PJM in April 2023 \cite{pjm_data} to determine the prices of the energy and regulation markets.\par
We generated extreme and typical frequency regulation scenarios for the bidding problem.
To capture frequency signal uncertainty, we partition the parameter space into intervals, each with an occurrence probability $\pi_{\omega}$.
Dividing (-1,1) into intervals of 0.1 and calculate the probabilities  for each. 
Extreme scenarios, when $s_{t,d}=$ -1 or 1, are handled separately. 

\begin{figure}[!ht]
    \centering  
    \vspace{-10pt}
    \subfloat[]{\includegraphics[width=0.25\textwidth]{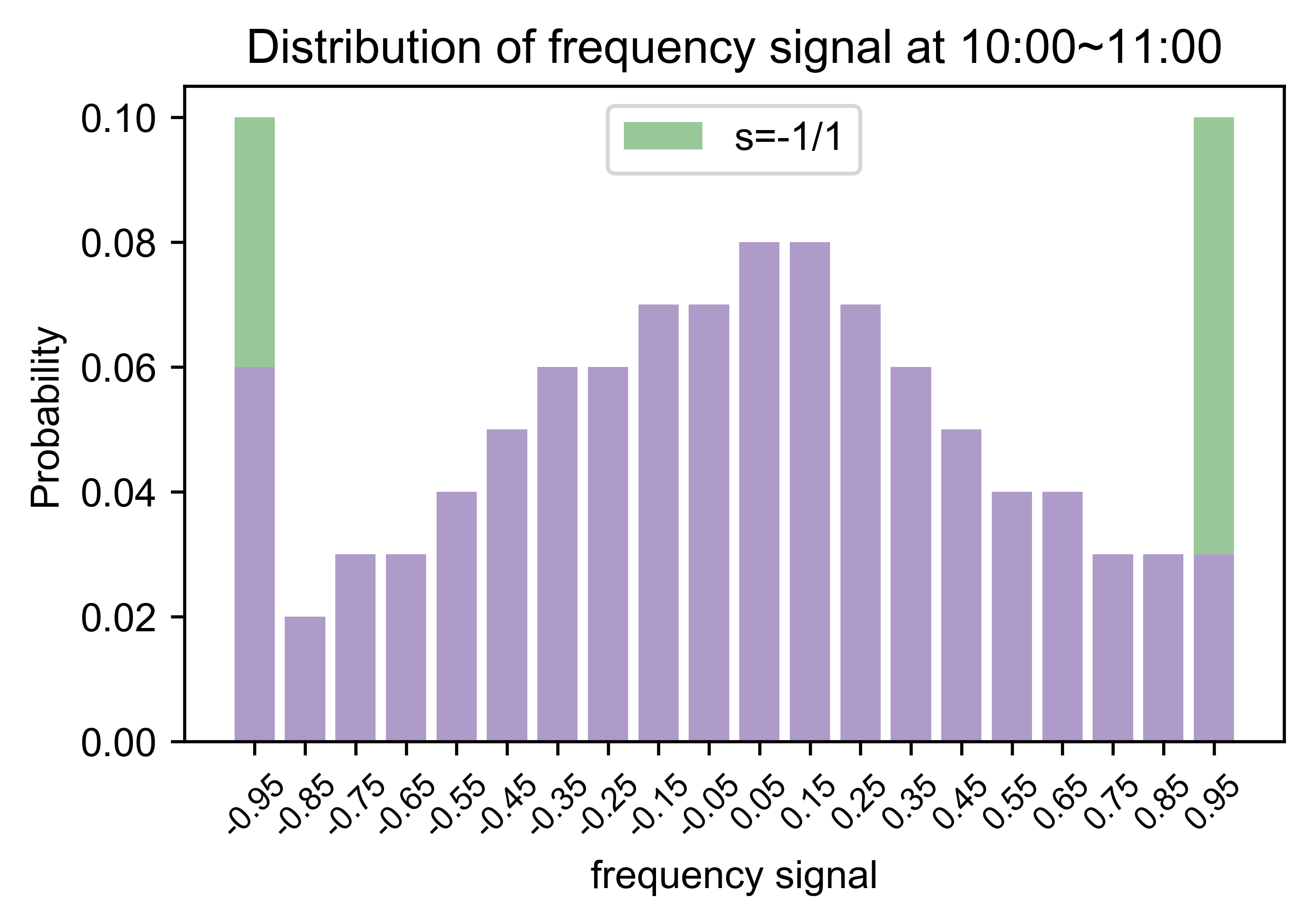}\label{sig_fig1}}%
    \subfloat[]{\includegraphics[width=0.25\textwidth]{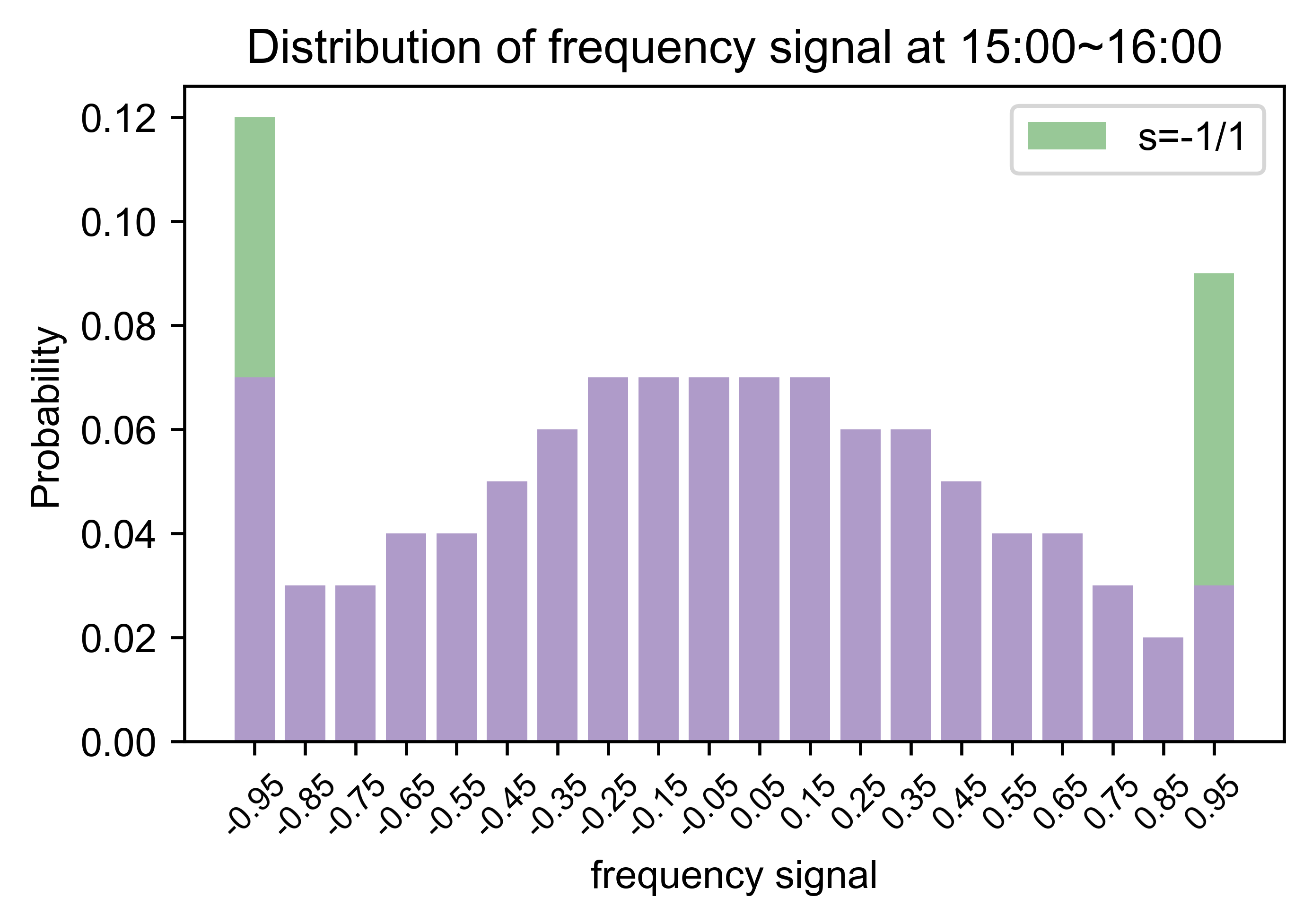}\label{sig_fig2}}%
    \captionsetup{font={footnotesize}, justification=raggedright}
	\caption{Distribution of regulation signal scenarios.}
    \label{fig:sig_dist}
\end{figure}

Using historical PJM RegD data in April 2020 \cite{pjm_data}, we calculate the average distribution of all scenarios, as illustrated in Fig. \ref{fig:sig_dist}. Extreme scenarios are represented by green bars, while typical scenarios are generally shown in purple bars, except for the boundary intervals. 

For example, Fig. \ref{sig_fig1} shows that the occurrence probabilities for $s_d=-1$ and $s_d=1$ are 0.041 and 0.069, respectively. For typical scenarios, the boundary scenarios [-1, -0.9) and [0.9, 1] have occurrence probabilities of 0.1 and 0.09, respectively. 
Additionally, each scenario uses the interval median value as the signal value. For example, for the interval [-0.9, -0.8), the corresponding signal value is $s_{t,d} = -0.85$.

\subsection{EV Flexibility Analysis}

\begin{figure}[ht]
    \centering
    \includegraphics[width=0.9\linewidth]{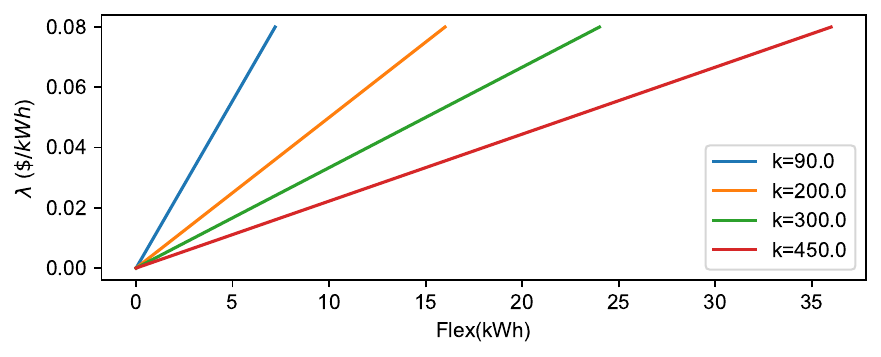}
    \caption{Flexibility bid-in supply curves in 21:00-22:00.}
    \label{fig:supply_curve}
\end{figure}

Based on the flexibility preference data shared by EV users, the flexibility supply curves are shown in Fig. \ref{fig:supply_curve}.



\begin{figure}[ht]
    \centering
    \subfloat[conservative]{\includegraphics[width=0.24\textwidth]{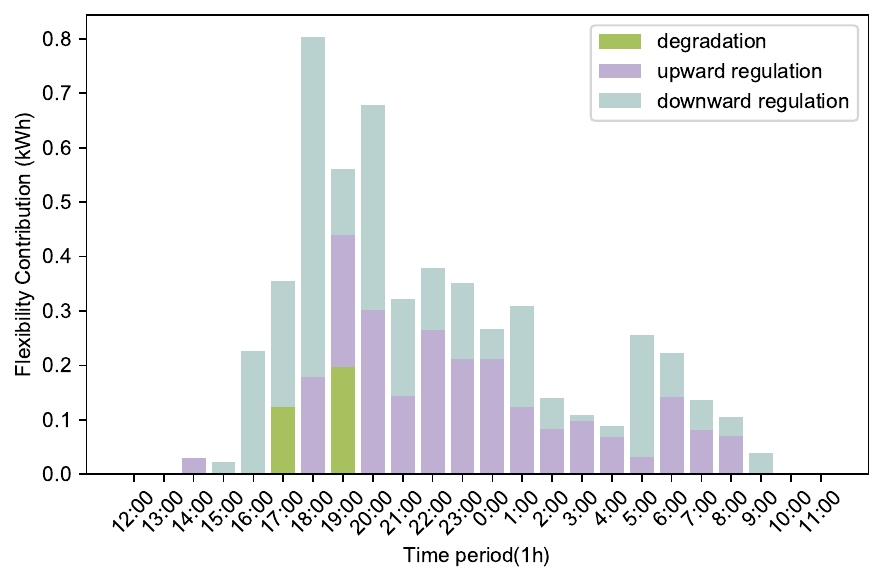}}%
    \label{group0}
    \hfil
    \subfloat[cautious]{\includegraphics[width=0.24\textwidth]{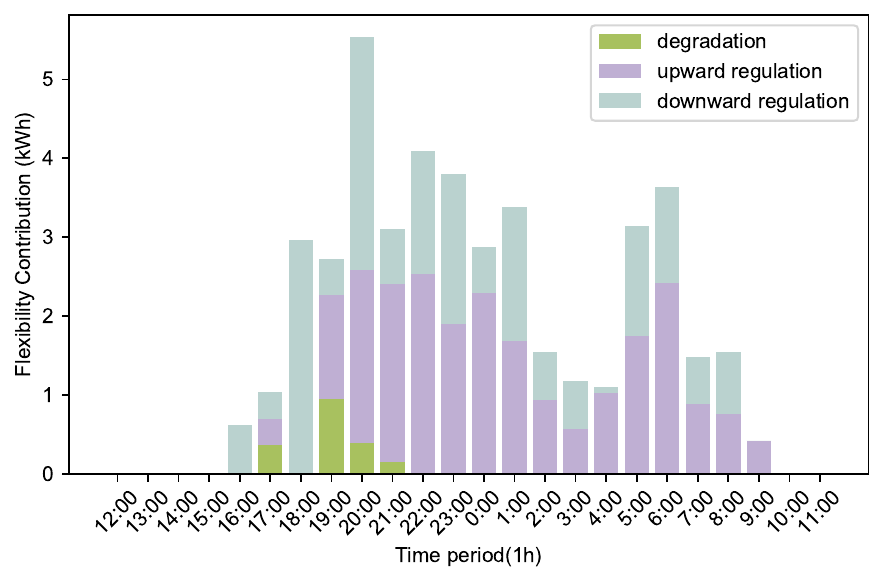}}%
    \label{group1}\\
    \vspace{-10pt}
    \hfil
	\subfloat[proactive]{\includegraphics[width=0.24\textwidth]{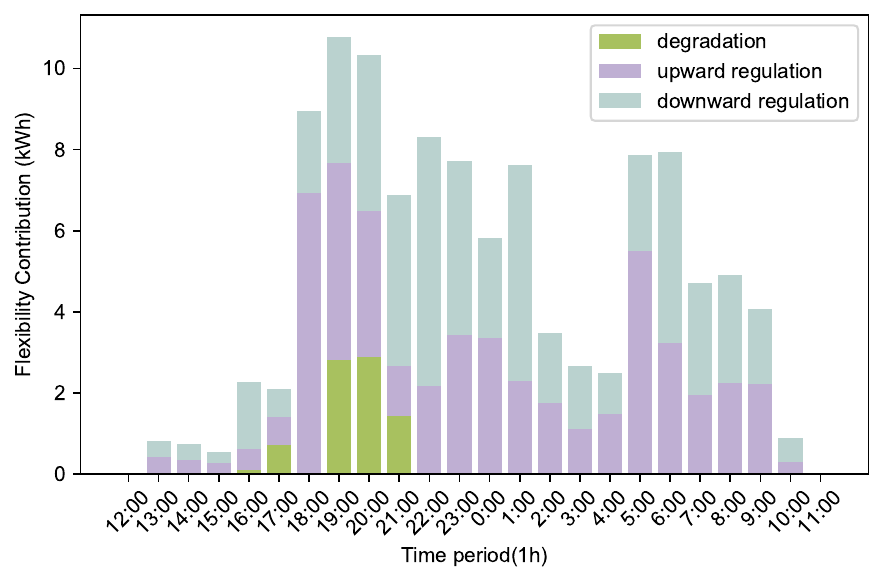}}
    \label{group3}
    \subfloat[risky]{\includegraphics[width=0.24\textwidth]{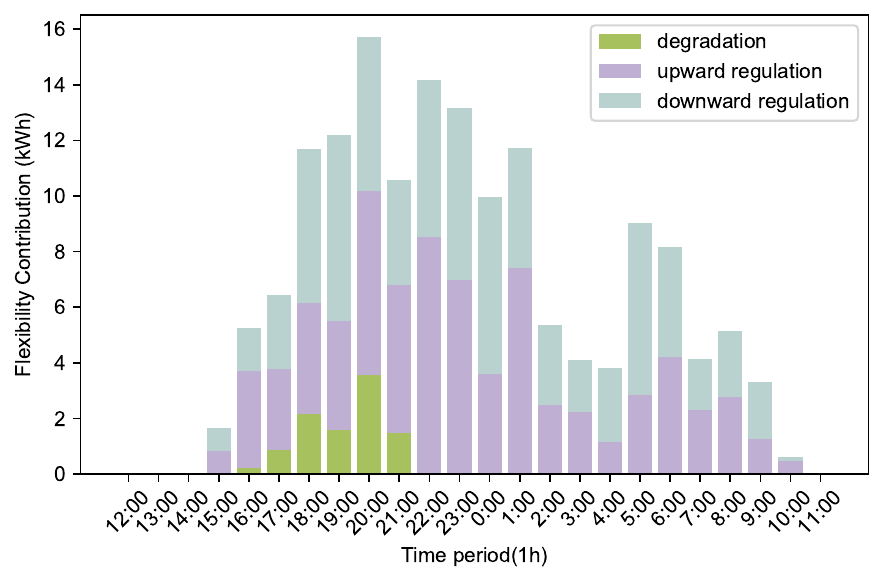}}
    \label{gurop4}
    \caption{Flexibility contribution of different EV groups.}
    \label{fig:flex_pre}
\end{figure}

Fig.\ref{fig:flex_pre} presents the total amounts of flexibility contributed by different types of EVs. It can be seen that with the increase of $k_n$, the flexibility contribution amount increases, where the risky EVs provide the most flexibility and the conservative EVs provide the least. This indicates that EVs with higher $k_n$ are more willing to contribute flexibility, which the EVA can leverage to obtain more flexibility from these users.\par
\begin{figure}[ht]
	\centering  
   \vspace{-10pt}
    \setlength{\abovecaptionskip}{0.cm}
    \subfloat[conservative]{\label{soc1}
     \includegraphics[width=0.24\textwidth]{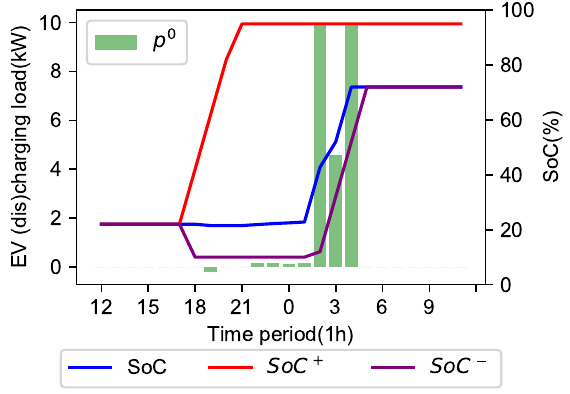}}%
    \subfloat[neutral]{\label{soc2}
    \includegraphics[width=0.24\textwidth]{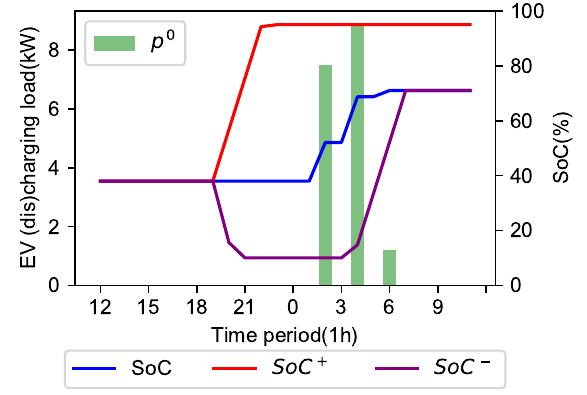}}%
    \vspace{-10pt}
	\subfloat[risky]{\includegraphics[width=0.24\textwidth]{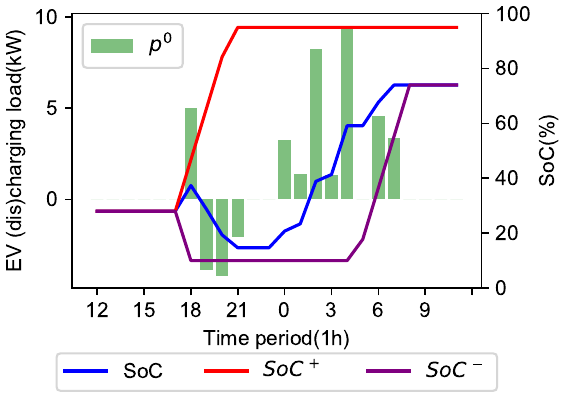}}%
    \label{soc3}
    \subfloat[risky]{\includegraphics[width=0.24\textwidth]{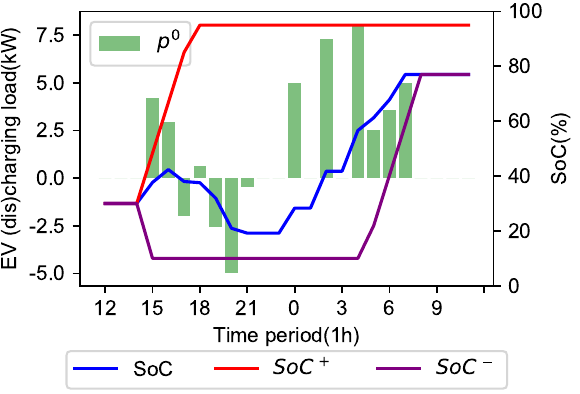}}%
    \label{soc4}
    \captionsetup{font={footnotesize}, justification=raggedright}
    \caption{Power and SoC curve of EV with different flexibility offering preference.}
    \label{fig:flex}
    \vspace{-20pt}
\end{figure}

The sub-figures in Fig.\ref{fig:flex} illustrate the energy dynamics of EVs with varying preferences for offering flexibility. The key difference arises from the EVA's cost-minimization strategy: 
the EVA can obtain flexibility from the risky EV with the smallest $k_n$ at a lower price. On the contrary, obtaining flexibility from conservative and neutral EVs triggers a higher cost for the EVA. To minimize its total cost, EVA tends to obtain flexibility from risky EVs and then starts to obtain that from conservative and neutral EVs after the flexible EVs have been fully exploited.

\subsection{EVA Bidding Results Analysis and Comparison}
\begin{figure}[!ht]
    \includegraphics[width=\linewidth]{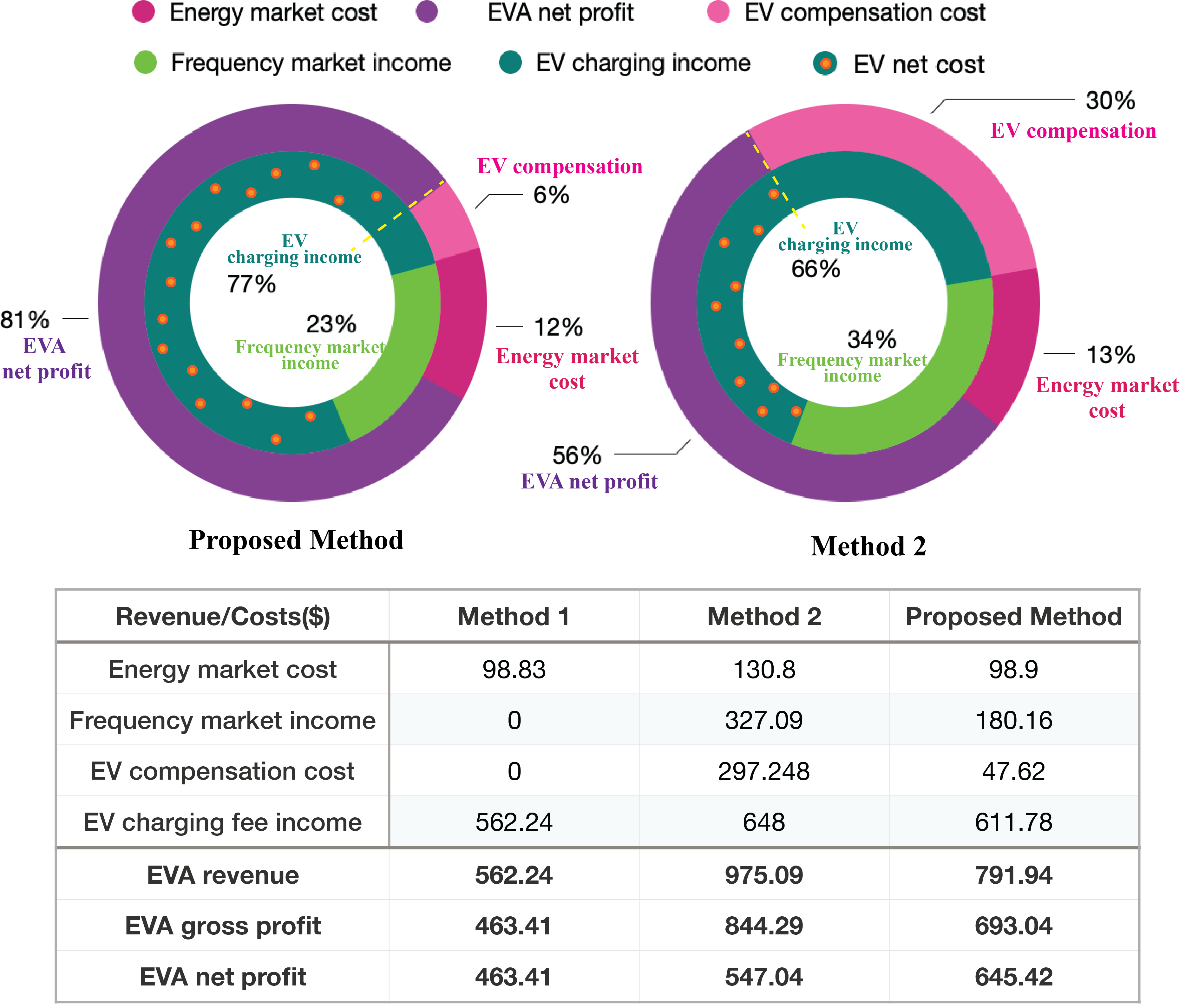}
    \caption{EVA Revenue and Costs.}
    \label{fig:bid_res2}
\end{figure}

\begin{figure*}[!ht]
    \centering  
    \subfloat[Method A]{\includegraphics[width=0.32\textwidth]{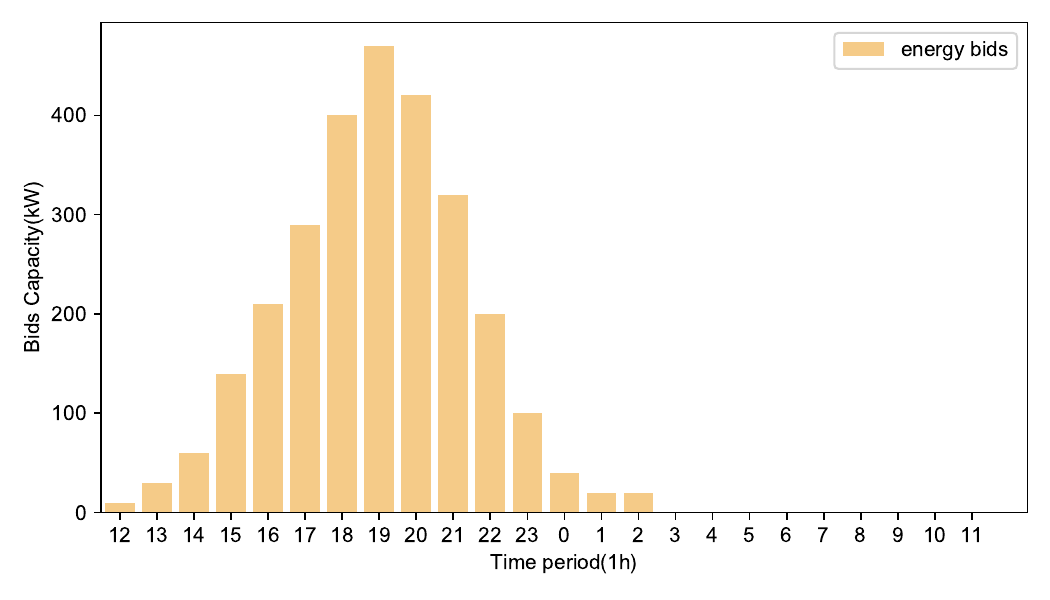}\label{Method_a}}
    \subfloat[Method B]{\includegraphics[width=0.32\textwidth]{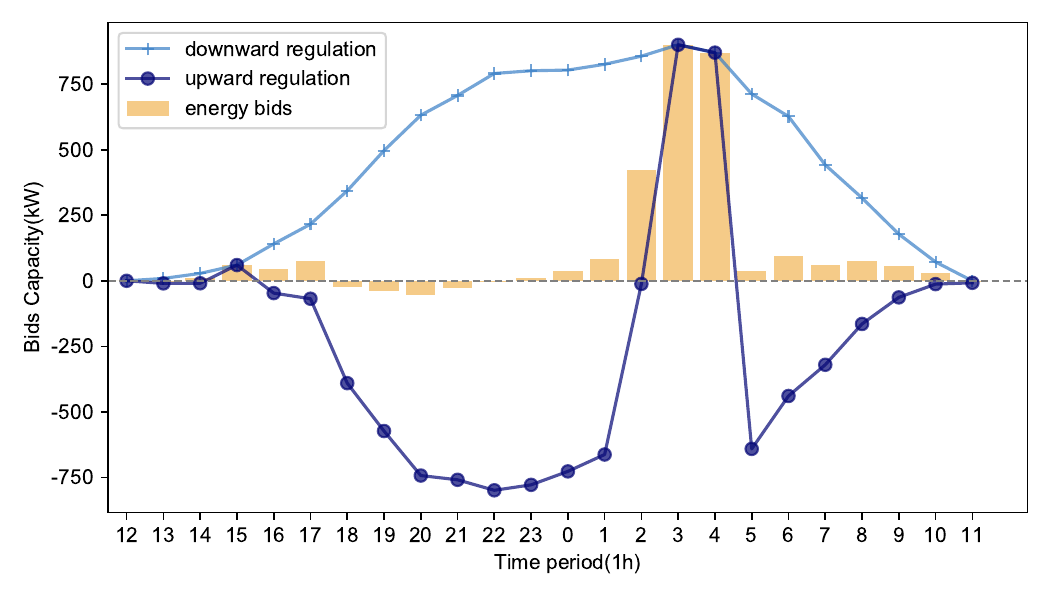}\label{Method_b}}
    \subfloat[Proposed method]{\includegraphics[width=0.32\textwidth]{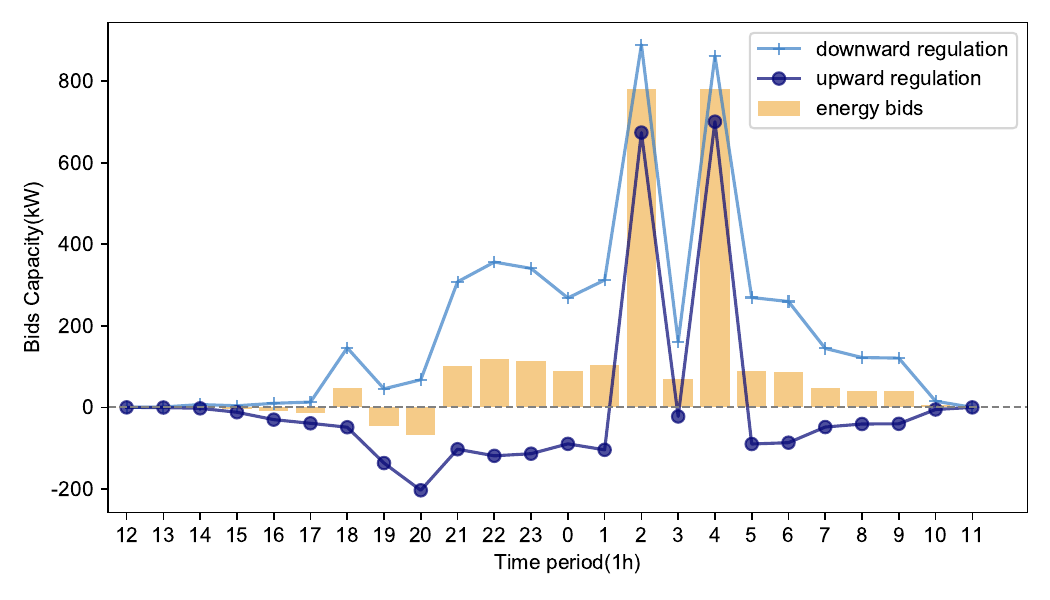}\label{Method_proposed}}%
    \captionsetup{font={footnotesize}, justification=raggedright}
	\caption{Bidding results of different methods
        (a) market prices,
        (b) Distribution of regulation signal scenarios at 22:00-23:00.}
            \label{fig:bid_res}
\end{figure*}

Two bidding strategies are used for comparison with the proposed bidding strategy.
Method 1 is to fulfill the EVs' charging demand as soon as possible, which is commonly adopted in practice.
Method 2 is an extension of the strategy presented in \cite{allocation3}, where the EVA aims to maximize its total profit while neglecting the flexibility offering preference of EVs. For comparison, we retrospectively assess the flexibility contributions of EVs and compensate them according to their flexibility supply curve.
Fig. \ref{fig:bid_res2} summarizes the revenue and costs of the EVA for each strategy, while the bid-in quantities are presented in Figure \ref{fig:bid_res}.

The proposed method balances the EVA’s profitability with the interests of EV users. 

In Method 1, the EVA fully respects EV users' charging demands, resulting in relatively low charging fees. According to the proposed definition of flexibility, this method will not offer any flexibility. However, the proposed method encourages EVs to offer flexibility by compensating for their contributions, allowing the EVA to participate in the frequency market.

In Method 2, the EVA benefits more in the frequency regulation market than the proposed method. This is because it assumes that the EVA can exploit as much flexibility from EVs as it desires, regardless of the willingness of the EV users.
However, this assumption is unrealistic for two reasons:
\romannumeral1) Offering flexibility relies on the rapid response characteristics of EVs and can potentially harm their batteries.
\romannumeral2) Providing flexibility results in a loss of opportunity costs, indicated by the rising energy market costs, compared to both Method 1 and the proposed method.
In this context, EVs will have no incentive to accept Method 2 to offer such substantial flexibility with even higher charging costs.
When we assess the flexibility contributions of EVs and compensate them according to their flexibility supply curve, the net profit for the EVA is lower than that of our proposed method.

\begin{figure}[ht]
  \begin{center}
  \includegraphics[width=0.5\textwidth]{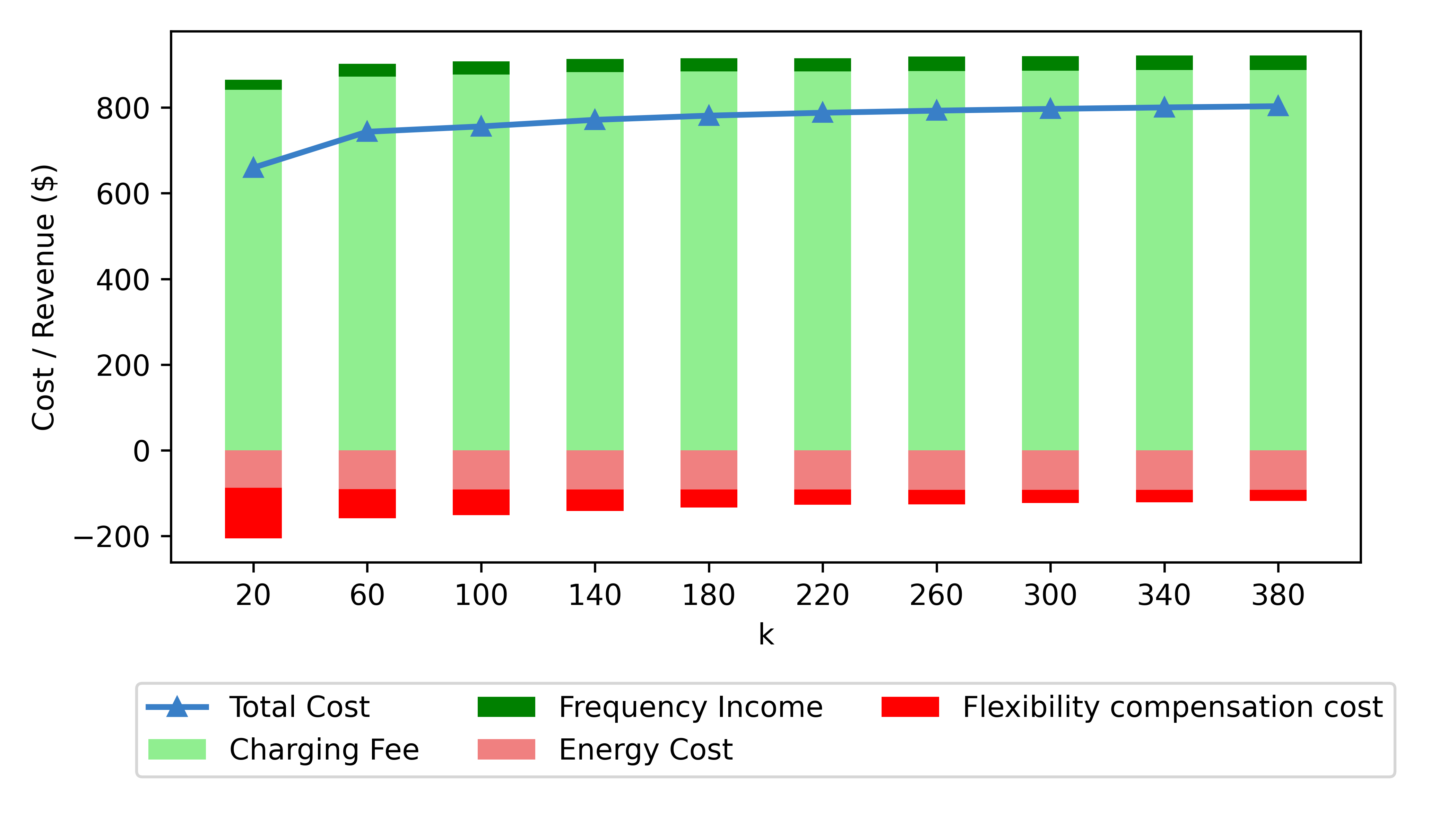}
  \caption{Trade-off between the profit of EVA and EV users’ willingness to offer flexibility} \label{fig_fair}
  \end{center}
\end{figure}

The trade-off between the EVA's profits and EV users’ willingness to offer flexibility is illustrated in Fig.\ref{fig_fair}.

Generally, EVA's overall profit increases when EV users are more willing to offer flexibility. However, this often results in a reduction in individual EV's profit. 
The balance between maximizing profits and maintaining fairness can be managed by adjusting the $k$ values.

\subsection{Power Dispatch Analysis}
In this subsection, we will evaluate the feasibility and fairness of the proposed power dispatch strategy.

\begin{figure}[ht]
    \centering
    \subfloat[optimal charging power]{\includegraphics[width=0.23\textwidth]{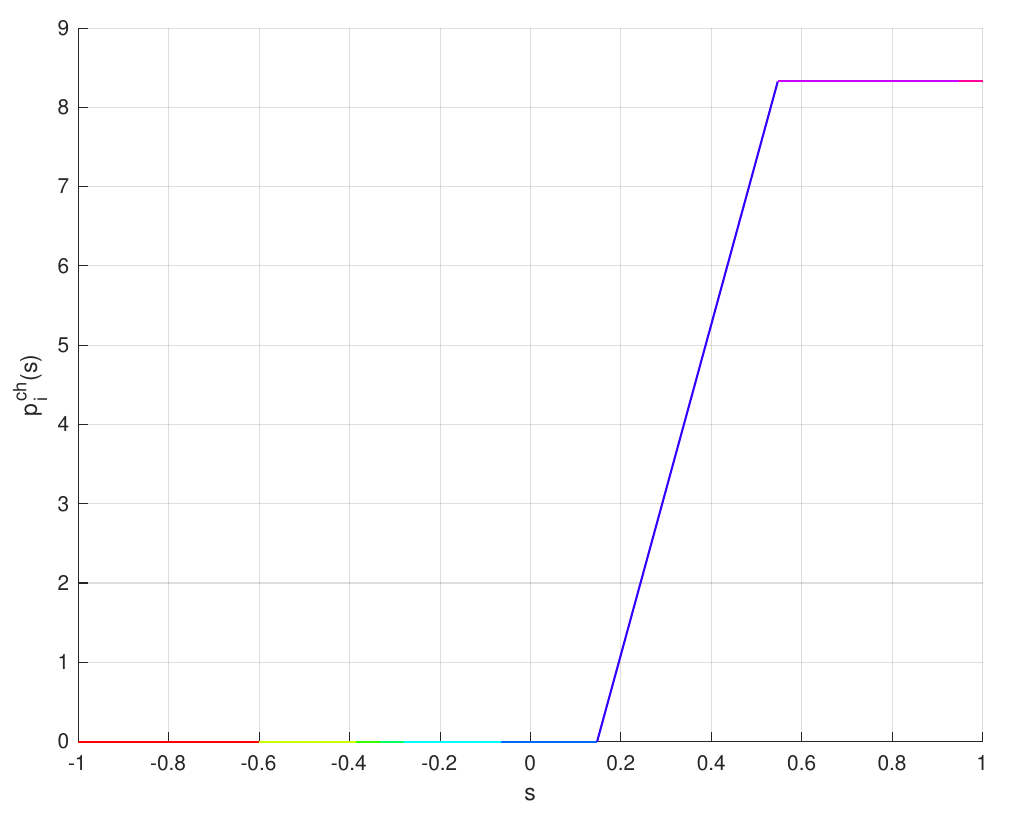}}\label{charging}
    \subfloat[optimal discharging power]{\includegraphics[width=0.23\textwidth]{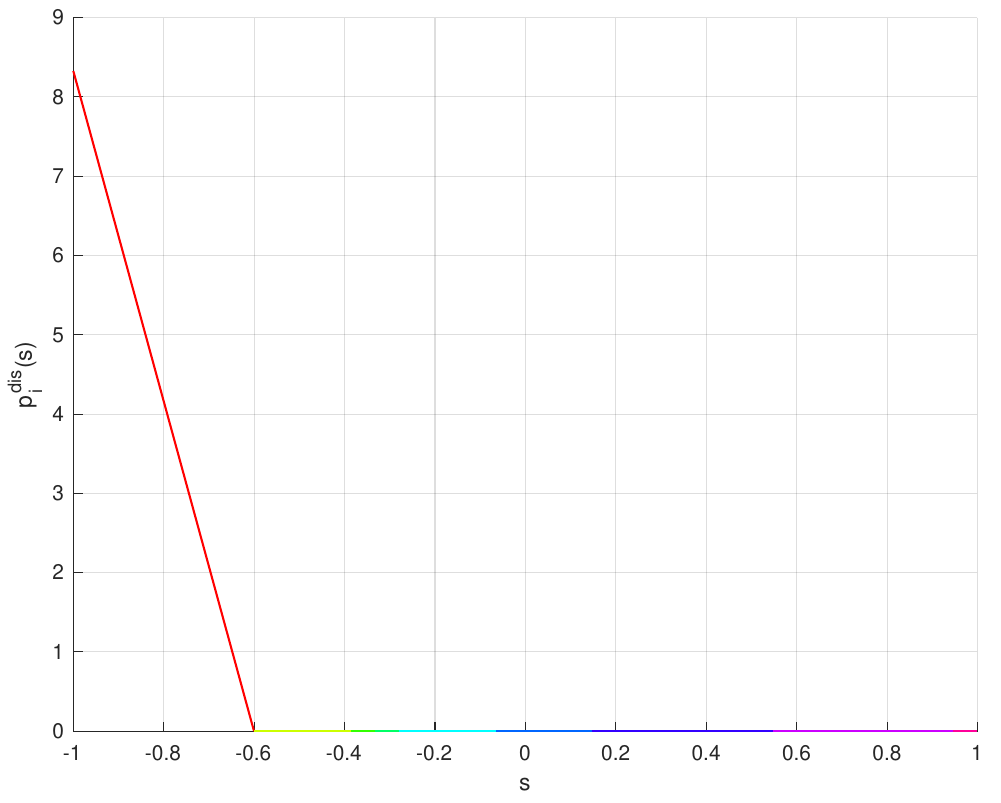}}\label{discharging}\\
    \subfloat[Value function]{\includegraphics[width=0.47 \textwidth]{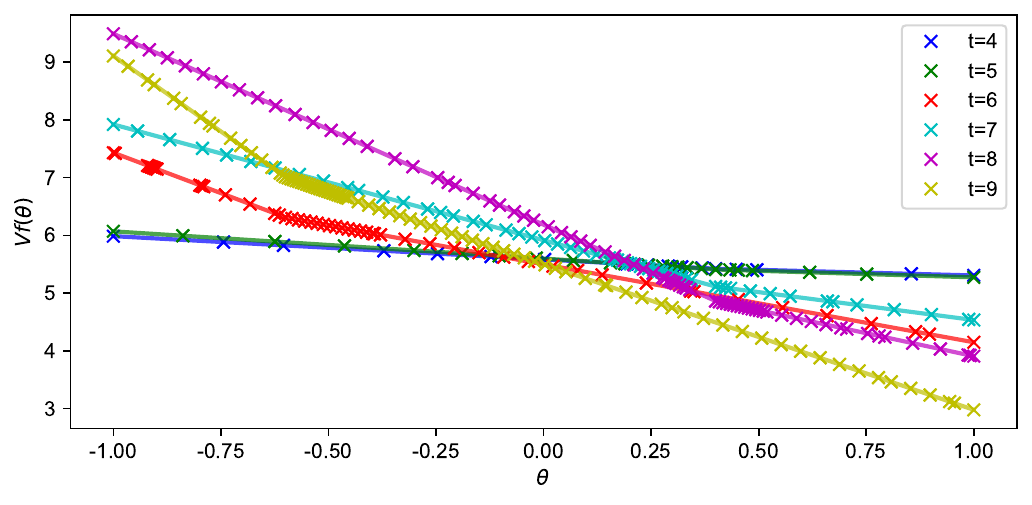}\label{affine}}\\
    \caption{Affine mapping of the dispatch problem}
    \label{fig:vf}
\end{figure}

For dispatch feasibility, the optimal dispatch cost and optimal dispatch decision can be determined based an the affine mapping optimal value function and solution function, both are affine  on the regulation signal, as shown in Fig. \ref{fig:vf}.

For dispatch fairness, we adopt Jain's fairness index \cite{fairness}, which respects five axioms: Continuity, Homogeneity, Asymptotic Saturation, Irrelevance of Partition, and Monotonicity. 
Jain's index is defined as follows:
\begin{align}
    J(\mathbf{x}) = \frac{\left( \sum_{i=1}^{N} x_i \right)^2}{N \sum_{i=1}^{N} x_i^2},
\end{align}
where $\mathbf{x} = (x_1, x_2, \dots, x_N)$ refers to the flexibility cost for each EV. Jain's index ranges from 0 to 1, with higher values indicating greater fairness.

The allocation methods compared are:
In \textit{proportional method}, power adjustments $\delta p ^{up/dn}$ are allocated proportional to the regulation contribution $\Delta p^{up/dn}$. In  \textit{round-robin method}, power adjustments $\delta p ^{up/dn}$ are distributed evenly to EV users with respect to their operational constraints. In \textit{maximum fairness method}, power adjustments $\delta p ^{up/dn}$ are allocated to maximize fairness.
The results are summarized in Table \ref{tab_disp}.

\begin{table}[h!]
  \begin{center}
    \caption{Dispatch results comparison}
    \begin{tabular}{c c c c } 
      \hline
      \textbf{Method} & \textbf{EVA costs (\$)}&\textbf{Fairness}& \\
      \hline
      Proportional   & 194.04 &  0.71   &    \\
      Round robin     & 178.03  & 0.53  &   \\
      Maximum fairness & 200.43  &  1   &     \\
      Proposed method & 149.61 & 0.82 & \\
      \hline
    \end{tabular}
    \label{tab_disp}
  \end{center}
\end{table}

The proportional method allocates power based on the regulation contribution but neglects the EV's flexibility preferences, resulting in high EVA costs. The round-robin method, while simple to implement, lacks fairness and results in relatively high EVA costs as well. The maximum fairness method achieves the highest fairness but at the expense of higher EVA costs.
The proposed method strikes a balance, achieving lower EVA costs while maintaining a high fairness index, making it a more practical solution.

\section{Conclusion}\label{s7}
This paper first proposes online EVA bidding and dispatch strategies for energy and frequency regulation that consider EV users' flexibility, which has not been properly addressed in both industry and academia.
It employs flexibility quantification and pricing in the EVA bidding model.
Furthermore, an affine mapping control strategy based on parametric linear programming is proposed to solve the power dispatch problem. 
Besides, numerical experiments have shown that the computational efficiency and profitability for EVA bidding and dispatch is acceptable. 
In practice, EV control may be limited by distribution network power flow constraints. Here, we assume these constraints are non-binding and can be ignored, as in \cite{flex3,reserve2,cvar,allocation3,zhanghongcai_regulation,reserve1}. Although power flow constraints are not explicitly included in our formulation, they can be incorporated if needed.
This paper focuses on EVA bidding and regulation capacity allocation, leaving power flow modeling for future work. Future research will also address unified DER flexibility aggregation and coordination across distribution and transmission levels.

\ifCLASSOPTIONcaptionsoff
  \newpage
\fi




\footnotesize
\bibliographystyle{IEEEtran}
\bibliography{IEEEabrv,Bibliography}

\vfill


\end{document}